\documentclass[a4paper]{article}
\usepackage[utf8]{inputenc}
\usepackage{url}
\usepackage{geometry}
\usepackage{graphicx}
\usepackage{hyperref}
\usepackage{biblatex}
\usepackage{amsmath}
\graphicspath{ {./notebooks/} }



\title{Factors determining maximum energy consumption of Bitcoin miners}
\author{Jesus M. Gonzalez-Barahona}
\date{July 19 2021, v0.9}

\begin{document}
\maketitle

\begin{abstract}
  \textit{Background:} During the last years, there has been a lot of discussion and estimations on the energy consumption of Bitcoin miners. However, most of the studies are focused on estimating energy consumption, not in exploring the factors that determine it.
  
  \textit{Goal:} To explore the factors that determine maximum energy consumption of Bitcoin miners. In particular, analyze the limits of energy consumption, and to which extent variations of the factors could produce its reduction. 

  \textit{Method:} Estimate the overall profit of all Bitcoin miners during a certain period of time, and the costs (including energy) that they face during that time, because of the mining activity. The underlying assumptions is that miners will only consume energy to mine Bitcoin if they have the expectation of profit, and at the same time they are competitive with respect of each other. Therefore, they will operate as a group in the point where profits balance expenditures.
  
  \textit{Results:} We show a basic equation that determines energy consumption based on some specific factors: minting, transaction fees, exchange rate, energy price, and amortization cost. We also define the Amortization Factor, which can be computed for mining devices based on their cost and energy consumption, helps to understand how the cost of equipment influences total energy consumption.

  \textit{Conclusions:} The factors driving energy consumption are identified, and from them, some ways in which Bitcoin energy consumption could be reduced are discussed. Some of these ways do not reduce the most important properties of Bitcoin, such as the chances of control of the aggregated hashpower, or the fundamentals of the proof of work mechanism. In general, the methods presented can help to predict energy consumption in different scenarios, based on factors that can be calculated from available data, or assumed in scenarios.
\end{abstract}

\section{Introduction}

Several studies have estimated the total energy consumption of all Bitcoin miners, starting at least in 2014~\cite{odwyer14:_bitcoin_minin_energ_footprint,mccook14:_bitcoin}, with later followups, updates and extensions. To some extent, the topic attracted the interest of the media in 2018, with some studies that compared Bitcoin mining with other activities, such as gold mining~\cite{Krause2018QuantificationOE}. Several observatories have been established, that claim to track the energy consumption of Bitcoin such as the Cambridge Bitcoin Electricity Consumption Index (CBECI)\footnote{Cambridge Bitcoin Electricity Consumption Index: \url{https://cbeci.org/}} and the Bitcoin Energy Consumption Index (BECI)\footnote{Bitcoin Energy Consumption Index: \url{https://digiconomist.net/bitcoin-energy-consumption}}.

When I was checking those studies and the methodology used in those observatories, I found that most of them used some variation of the same general method. As data sources, they use an estimation of the the total hashpower of all Bitcoin miners, and some assumption about the equipment used by miners (in terms of hashpower and energy consumption). To compute the total energy consumption, they compute how much equipment would be needed to produce the total hashpower, and then add up the energy consumption of all that equipment.

In this paper, I'm using a method which is not novel, but is less used in those studies (it is used for example in the BECI). It starts with an economic assumption: miners would only mine Bitcoin if it makes economical sense for them, that is, when they found economical benefit in it. In other words, the total costs faced by miners would be lower then their expected income. Therefore, since energy cost is one among other costs that miners face, energy costs for all miners together would not be higher than expected income by all miners together. In fact, if other costs (such as amortization of equipment) could be estimated, upper bounds on energy consumption could be more precisely estimated as well. Fortunately, in the case of Bitcoin the expected income is a relatively easy function, as will be shown below, which makes the estimation of this upper bound feasible.

Exploring this assumption was useful for me to better understand the factors that determine Bitcoin energy consumption, and how they interact. This exploration led me to some interesting consequences, including:

\begin{itemize}
\item The aggregated hashpower of all Bitcoin miners together has no impact on the upper limit for their aggregated energy consumption. In fact, aggregated hashpower is the consequence of other factors.
\item The exchange rate of Bitcoin to traditional currencies has a linear impact on the upper bound for the aggregated energy consumption: the higher the rate, the higher the upper bound for energy consumption.
\item Amortization costs have a negative impact on the upper bound of energy consumption: the higher the cost of the equipment for mining, the lower the upper bound.
\item Energy price has a negative impact on the upper bound of energy consumption: the higher the energy price, the lower the upper bound.
\item A single number (the Amortization Factor) can be defined to characterize mining equipment. This single number drives the split in costs for the miners between amortization of equipment and energy consumption.
\item There are some actions that could lead to reducing the aggregate energy consumption by Bitcoin miners without affecting the main properties of Bitcoin.
\end{itemize}

Some of those consequences are found only if some other assumptions hold, which is something that could be the subject of further research. Those assumptions will be presented in detail later in this paper. On the other hand, some of these consequences may also seem obvious once you think about them. For example, the fact that if energy price increases and all other factors remain stable miners will adapt by consuming less energy, seems straightforward from a maximization of benefits point of view. However, showing the relationship with equations helps to explore it in detail.

The method is based on three main factors (price of energy, exchange rate of Bitcoin to traditional currencies, and amortization cost for equipment), and to some extent on a fourth one (amount of fees in Bitcoin transactions), and maybe a fifth one (minted coins per block, which only changes every four years). Using this method, different strategies for lowering the upper bound of energy consumption can be explored, or the impact on energy consumption of the exchange rate reaching some value can be estimated.

The rest of this paper is structured as follows. First, we present the Bitcoin protocol (how Bitcoin work, and how Bitcoin ``miners'' mine blocks), and some definitions, in the next section. Then, Section~\ref{sec:income-cost} presents the analysis of how miners get their income, and how they spend it to mine Bitcoin. Section~\ref{sec:energy-consumed} shows how the energy consumed by all miners can be computed based on the analysis of income and expenses, and proposes an equation showing which factors affect energy consumption. Section~\ref{sec:current-energy} computes the maximum energy consumed in some scenarios (some of them maximalist or minimalist, and a medium scenario roughly similar to the situation in June 2021). Then, Section~\ref{sec:impact-factors} explores in some detail the impact of the different factors on energy consumption. The paper ends with some discussion on the results, including some proposals for actions that could impact energy consumption (Section~\ref{sec:discussion}), a brief discussion of related work (Section~\ref{sec:related-work}), and some conclusions (Section~\ref{sec:conclusions}).

Please, while reading through the paper remember that I'm making no claim that the approach is novel, neither even that it is correct. Any feedback about prior work, about errors in the approach will be more than welcome, or about anything else, is welcome. From that point of view, please consider this paper as a request for comments.

\section{The Bitcoin protocol}

There are many descriptions of how Bitcoin works, both from a technical and an economic point of view. From the point of view of the knowledge useful for following this paper, maybe the read can refer to~\cite{bohme15:_bitcoin}, in case of not being familiar with the concepts presented here. Of course, the original paper by Satoshi Nakamoto will also be helpful~\cite{nakamoto09:_bitcoin}. A detailed discussion on Bitcoin characteristics, mainly from an economic point of view, but also detailing how their properties emerge from the Bitcoin protocol can be found in~\cite{chen20:blockchain_economics}.

This paper is exclusively focused on Bitcoin, and although the approach and some of the conclusions could be extended to other blockchain-based currencies, there is no exploration of that in it.

For the purposes of introducing the main characteristics of it relevant for this paper, we can assume Bitcoin works as follows. First of all, Bitcoin is implemented as a persistent, distributed ledger. The unit of information in that ledger is a transaction, and all transactions are ordered in the ledger. Each transaction codifies how a number of Bitcoins change hands (or to be more precise, change from one or more public keys to one or more public keys, which all public keys being associated to Bitcoin wallets).

Transactions are grouped in blocks, and a block is added to the ledger every approximately 10 minutes, forming the Bitcoin blockchain. Adding a block to the blockchain is a key procedure. At any given moment, a lot of computers are competing for adding a block, with a certain collection of transactions. The Bitcoin protocol guarantees that only one of them will eventually win, thus maintaining a single blockchain. Computers compete by performing computations, in which is named as ``proof of work''. When a computer ``wins'', and adds a block to the blockchain, it gets in that block a transaction with some new Bitcoin for a wallet of its preference, in a process which is named ``minting''. This minting process is how Bitcoin are created. The number of Bitcoin that are obtained when a block is added to the blockchain is halved every approximately 4 years, so that the total number of Bitcoin that can be ever minted is fixed and known in advance.

Transactions can also specify a fee. The fee is decided by whomever does the transaction, and when a computer builds a block to try to add it to the blockchain, it decides which transactions to include it, and usually tries to maximize profit by including transactions with a higher fee. The reason is that the block includes a transaction collecting fees, so the computer whose block is chained decides to which wallet are sent the fees for the transactions in it. This usually means that the larger the fee in a transaction, the most likely that it is quickly included in the blockchain.

In the rest of this paper, we refer to ``miner'' as a computer trying to include blocks in the blockchain, a process which is usually named ``mining''. A ``miner's wallet'' will be a wallet controlled by whomever is running the mining operations on that miner, which can be used to collect minted Bitcoin or transaction fees. We refer to ``block'' as any of the elements of the Bitcoin blockchain, including a set of transactions. When the block is inserted in the blockchain, those transactions are inserted into it too, and after a few more blocks are inserted, are considered to be immutable. We will refer to the operation of inserting a new block in the blockchain as ``mining a block''.

We will refer to ``hashpower'' (in some cases ``hash rate'') as the number of hashes per unit of time that can be computed by some miner (or by all miners aggregated). The number of hashes captures how fast are performed the operations to mine a new block, and when referred to the whole set of miners, reflects how ``difficult'' it is to mine a block: the more hashpower, the more difficulty for mining a block.

Even when these definitions are not rigorous, we consider they are clear enough in the context we use them in this paper.

\section{Income and costs for all miners}
\label{sec:income-cost}

The basis of the analysis is the balance of income and expenses of the whole set of miners mining blocks for Bitcoin. Therefore, we will start by analyzing income and expenses for the set of all Bitcoin miners. Instead of analyzing income and expenses for each miner, and then add all of them together we will analyze from the beginning all the set of miners.

\subsection{Income}

Miners have two economic incentives for mining, both received in Bitcoin, both received when they chain a new block to the Bitcoin blockchain (they ``mine'' a block). So, both incentives are income per block. They were briefly explained above, when the functioning of Bitcoin was introduced, and they are:

\begin{itemize}
\item Minting. Each time a miner mines a new block some Bitcoins are allocated to the miner's wallet. The number of Bitcoins that are minted for each block is reduced by half every 210,000 blocks (or about four years), being currently (2021) 6.25 Bitcoins per block.
\item Fees. Each transaction can propose a fee, and it usually does. The miner that mines a block with that transaction can transfer the fee to a wallet under its control. Therefore, the miner that mines a block ``collects'' the fees in all the transactions in that block. Each transaction can include a different fee, and fees usually vary depending on several factors, the number of pending transactions being one of the most important.
\end{itemize}

Therefore, if we consider $Miners$ as the set of all the miners working over a certain time period ($T$), the total income collected by $Miners$ during $T$ is:

\begin{equation}
  {Income}_T = ({Blocks}_T \times {Mint}) + {Fees}_T
  \label{eq:income}
\end{equation}

with the following definitions:

\begin{itemize}
\item $Income_T$: Total income by $Miners$ during the period $T$, in Bitcoin.
\item $Blocks_T$: Number of blocks mined by $Miners$ during the period $T$. Since $Miners$ includes all miners working during the period, this is equal to the total number of blocks mined during that period.
  
\item $Mint$: Income due to minting for a single block, in Bitcoin, during the period $T$. For simplicity, this is assumed to be constant during the period $T$.

\item $Fees$: Sum of all fees collected from all the transactions in blocks mined by $Miners$ during the period $T$. Again, since $Miners$ include all miners mining blocks, and therefore, all miners collecting fees, this is equal to all fees collected during $T$, and to all fees in all transactions in blocks mined during $T$.
\end{itemize}

It is important to notice that total income, in Bitcoin, is not related to the total hashpower of all the miners, is predictably related to the period considered, and is sensible to variations in fees.

Total income by all miners is not related to the total hashpower because whatever it is, the number of Bitcoin collected by all miners is the same: it will only be more or less difficult to obtain in aggregate for all of them. This is due to how the total hashpower is adjusted, approximately every week, to keep the block rate approximately constant in one block about every ten minutes.

Income is predictable related to the period considered ($T$), because (${Blocks}_T\times{Mint}$ is known in advance, given that $Blocks$ will be approximately equal to the number of 10 minutes periods in $T$.

Income is sensible to variations in fees, since fees are a sizable part of the income, and they are known to vary over time. In fact, since the fraction of income due to minting will decrease over time: $Mint$ is halved every four years, which means that ${Blocks}_T*{Mint}$ will be much smaller in some years.

\subsection{Expenses for all miners}

Expenses needed to keep miners in production can be split in:

\begin{itemize}
\item Expenses due to the hardware being used to mine, and all the related investment to keep it housed and in good condition. Although these expenses may be in a large part a capital expenditure, we will assume its amortization over a period $T$ can be calculated.
\item Expenses due to operation expenditure. A good part of current literature assumes these costs are driven almost exclusively by power consumption.
\end{itemize}

For the rest of this paper, we will assume all costs (expenses) of a miner are for a certain period of time $T$:

\begin{equation}
  Costs_T = Amort_T + (Energy_T \times EnergyPrice)
\end{equation}

with the following definitions:

\begin{itemize}
\item $Costs_T$. All costs (expenses) for $Miners$ over the period $T$, in Bitcoin.
\item $Amort_T$. All amortization costs for $Miners$, calculated for the period $T$, in Bitcoin. In practice, we include in this term all costs that are not due to energy consumption for operation, and assume it can be computed for a given period.
\item $Energy_T$. All energy consumed by $Miners$ for operation during the period $T$, in KWh (KiloWatt-hour).
\item $EnergyPrice$. Price of energy during period $T$. For simplicity, it is assumed that the price is constant during $T$, and equal for all miners. In practice,  $EnergyPrice$ could be calculated as the weighted average of the price of energy for all miners (weighted by the amount of energy consumed).
\end{itemize}

It is important to notice that costs are not related to the total hashpower of $Miners$. Of course, if more hashpower is used to produce blocks, more energy will be needed, for a given kind of hardware. But in any case, this is due to the number and quality of miners joining $Miners$, not to the hashpower needed, which will be adjusted up or down depending on the aggregated hashpower of $Miners$ grow or shrink.

\section{Energy that miners can consume for mining}
\label{sec:energy-consumed}

Assuming miners act as rational economic actors, they will operate when their mining costs are below their expectations of income. To understand how much they can allocate to energy cost, we already analyzed income and costs. Therefore, miners will operate during a period $T$ trying to make the following inequality hold:

\begin{equation}
  Costs_T \leq Income_T 
\end{equation}

or

\begin{equation}
  Amort_T + (Energy_T \times EnergyPrice) \leq
  (Blocks_T \times Mint) + Fees_T
  \label{eq:cost-income-equilibrium}
\end{equation}

Therefore, we can calculate an upper limit on the energy consumed by $Miners$, which is the energy consumed by all Bitcoin miners, as:

\begin{equation}
    Energy_T \leq \frac{(Blocks_T \times Mint) + Fees_T - Amort_T}{EnergyPrice}
\end{equation}

This means that, for a given period $T$, the maximum energy consumed by all Bitcoin miners (if they don't operate under expectation of losses), $EnergyMax_T$ is:

\begin{equation}
    EnergyMax_T = \frac{(Blocks_T \times Mint) + Fees_T - Amort_T}{EnergyPrice}
\end{equation}

Up to this point, income and cost have been computed in Bitcoin. However, those costs are usually converted to US Dollar, Euro, or some other traditional currency. Since the price of Bitcoin with respect to those currencies has a high volatility, it is interesting to include that conversion ratio in the equation:

\begin{equation}
  EnergyMax_T = \frac{
    (Blocks_T \times \frac{MintEur}{EurBTC}) +
    \frac{FeesEur_T}{EurBTC} -
    \frac{AmortEur_T}{EurBTC}}
    {\frac{EnergyPriceEur}{EurBTC}}
\end{equation}

with the following definitions:

\begin{itemize}
\item $EurBTC$: exchange rate in Euros per Bitcoin.
\item $MintEur = Mint \times EurBTC$
\item $FeesEur_T = Fees_T \times EurBTC$
\item $AmortEur_T = Amort_T \times EurBTC$
\item $EnergyPriceEur = EnergyPrice \times EurBTC$
\end{itemize}



This equation can be transformed easily in the following one, which we call the ``Bitcoin Energy Factors Formula'':

\begin{equation}
  {EnergyMax}_T =
    \frac{(Blocks_T \times Mint \times EurBTC)} {EnergyPriceEur} +
    \frac{Fees_T \times EurBTC} {EnergyPriceEur} -
    \frac{AmortEur_T} {EnergyPriceEur}
    \label{eq:energy-factors}
\end{equation}

In this last equation, we have tried to keep each cost or income on the most usual currency (Bitcoin or Euro), according to how people tend to interact with it. For example, when people buy hardware or pay for housing costs ($AmortEur$), or pay for electricity ($EnergyPriceEur$), usually they pay for it in Euro (being Euro just a proxy for all ``traditional'' currencies). However, when they get minted coins ($Mint$), they get Bitcoin. For fees ($Fees$), we have used also Bitcoin as the unit, since people usually express fees in that currency when specifying Bitcoin transactions. To make the equation fit, we have used the exchange ratio of Bitcoin to Euro (EurBTC) whenever convenient. In any case, all of this is only to make the following discussion more easy to follow: results should be the same if you use other currencies for each factor.

Since this last equation shows that the maximum amount of energy that all Bitcoin miners can consume during a period $T$, while still having some expectation of profit, we can see that it depends on $Blocks_T$, $Mint$, $Fees_T$, $AmortEur$, $EurBTC$ and $EnergyPriceEur$. And nothing else.

The number of blocks per period of time ($Blocks_T$) is (approximately) known in advance, and does not change. The Bitcoin protocol adjusts every week to produce a block about every 10 minutes. Therefore, except for (unforeseen for now) changes in the core Bitcoin protocol, there is no influence of this factor in energy consumption, since it remains invariable. $Mint$ is invariable during long periods of time (about four years). Therefore, the only factors that influence the maximum energy consumption for all miners are fees collected from transactions, amortization costs, the exchange rate of Bitcoin, and the price of energy.

\section{Maximum energy consumed in some scenarios}
\label{sec:current-energy}

\begin{table}
  \begin{tabular}{|r||r|r|r|r|}
    \hline
    Name & $Fees_{day}$ & $EurBTC$ & $AmortEur_{year}$ & $EnergyPriceEur$ \\
         & (BTC)    & (EUR)  & (MEUR)        & (EUR) \\ \hline\hline
    Low income/low cost & 30.00 & 10,000 & 0 & 0.01\\
Low income/high cost & 30.00 & 10,000 & 10,000 & 0.10\\
Medium income/medium cost & 100.00 & 30,000 & 5,000 & 0.03\\
High income/low cost & 300.00 & 100,000 & 0 & 0.01\\
High income/high cost & 300.00 & 100,000 & 10,000 & 0.10\\

    \hline
  \end{tabular}
  \caption[Scenarios]{Some scenarios for illustrating the different factors influencing maximum energy consumption by all miners. Low (high) income scenarios assume low (high) fees per transaction and low (high) Bitcoin exchange rate. Low (high) cost scenarios assume low (high) amortization costs, and low (high) energy price. All of them compared with the current situation, as of mid-June 2021 (which is similar to the medium scenario). For all scenarios, the parameters are: transaction fees per day, collected by all miners; exchange rate for Bitcoin to Euro; amortization costs for all miners, per year; and energy price (weighted average).}
  \label{table:scenarios1}
\end{table}

Let's define some scenarios in Table~\ref{table:scenarios1}, which we will use to illustrate the results of applying the previous equations. They present four extreme situations, and one which could be close to the real situation while writing this paper. But all of them are only illustrative. Each scenario is defined by fixing the values for some of the variables in Equation~\ref{eq:energy-factors}. Other factors will remain constant:

  \begin{description}
  \item{$Blocks_T$:} 52560 blocks
    ($6 blocks/hour \times 8,760 hours/year$, for a year of 365 days)
  \item{$Mint$:} 6.25 Bitcoin/block (halving in late 2024)
  \end{description}

Some rationale for parameters used in these scenarios:

\begin{description}
  \item{$Fees$} For the medium scenario, we used 100 Bitcoin/day, which is close to the values during June 2021, obtained from BTC.com\footnote{\url{https://btc.com/stats/fee}}. This value was close to 500 Bitcoin/day in December 2017, but has remained below 130 Bitcoin/day since February 2018. For a low income scenario, we used 30 Bitcoin/day, way below fees during the last years, and for high income, 300 Bitcoin/day, lower than historical fees, but higher than fees during the last years.

  \item{$EurBTC$} We consider a range of 10,000 (low income) to 200,000 (high income) Euro to Bitcoin exchange rate, with a medium scenario of 30,000, which is close to current exchange. Of course, these values are arbitrary, but try to reflect extremely low exchange and extremely high rates, by today expectations.

  \item{$AmortEur_{year}$} For aggregated amortization we have considered an extreme scenario of 0 euros for low income, to show the case when miners won't care about amortization, either because they consider their equipment as amortized, or because it considers its cost as a sunken cost. For the medium scenario, we have considered 5,000 MEuro per year, close to the actual values calculated for June 2021. High cost scenarios use twice this value.

    To estimate the aggregated amortization costs of all miners we have computed the cost of the number of devices needed to produce the aggregated hashpower of all of them. We assume that devices used for mining are similar to some of the most profitable device models available in June 2021 (see~Table~\ref{table:devices}). For a period T, we can compute amortization costs for all miners ($AmortEur_T$) as:

\begin{equation}
  AmortEur_T = DeviceCostEur \times \frac{HashPower}{HashPowerDevice} \times T 
\end{equation}

being $DeviceCostEur$ the average cost of mining devices, $HashPowerDevice$ the average hashing power of mining devices, and $HashPower$ the aggregated hashing power of all active Bitcoin miners (in both cases, in THash/sec).

\begin{table}
  \begin{tabular}{|r||r|r|r|r|}
    \hline
    Name & $DeviceHashPower$ & $DevicePower$ & $DeviceEur$ & $AmortEur_{year}$ \\
         & (TeraHash/sec)       & (KW)            & (EUR)       & (MEUR)         \\ \hline\hline
    Antminer S19j Pro (100 Th) & 100.0 & 3.05 & 7,080 & 4,825.52 \\
Antminer S19j Pro (110 Th) & 110.0 & 3.25 & 8,797 & 5,450.89 \\
Antminer S9 SE (16Th) & 16.0 & 1.28 & 829 & 3,532.16 \\

    \hline
  \end{tabular}
  \caption[Hashing devices]{Hashing devices. List of some of the most profitable hashing devices available for purchase on June 13th 2021. Performance, consumed power, price, and estimated amortization cost for all miners, should all of them use this equipment. For estimating amortization cost, we use a total hash power of all of the miners of 136.316 ExaHash/sec. ($136.316 \times 10^6$ TeraHash/sec), which is the estimated total hash power of all miners as of June 13th 2021, according to Blockchain.com\protect\footnotemark. We consider a full amortization period of 2 years.}
  \label{table:devices}
\end{table}
\footnotetext{Blockchain.com (total hash rate)): \url{https://www.blockchain.com/charts/hash-rate}}

  \item{$EnergyPriceEur$} Energy price is difficult to estimate, since it should be the weighted average value of energy for all miners (weighted by their energy consumption). After looking at several references we have used for the medium scenario the value of 0.03 Euro/KWh based on the average price estimated by the World Bank in 2019\footnote{World Bank DB16-20 methodology, prices in US cents per KWh \\ \url{https://govdata360.worldbank.org/indicators/h6779690b}} for some countries like China (12.80 US cents per KWh), Iceland (12.20 US cents per KWh), Argentina (10.80 US cents per KWh), Saudi Arabia (7.40 US cents per KWh), and considering that usually miners tend to use electricity rates lower than the average in their countries. For the high cost scenario we used 0.10 Euro/KWh, which is probably higher than any price actually paid by miners in any country. For low cost, we used 0.01 Euro/KWh, which is likely lower than the real cost available to most miners.
  \end{description}

\begin{table}
  \begin{tabular}{|r||r|r|r|r|r|}
    \hline
    Name & $Fees_{day}$ & $EurBTC$ & $AmortEur_{year}$ & $EnergyPriceEur$ & $EnergyMax_{year}$ \\
         & (BTC)    & (EUR)  & (MEUR)        & (EUR)  & (TWh) \\ \hline\hline
    Low income/low cost & 30.00 & 10,000 & 0 & 0.01 & 339.45\\
Low income/high cost & 30.00 & 10,000 & 10,000 & 0.10 & -66.06\\
Medium income/medium cost & 100.00 & 30,000 & 5,000 & 0.03 & 198.33\\
High income/low cost & 300.00 & 100,000 & 0 & 0.01 & 4380.00\\
High income/high cost & 300.00 & 100,000 & 10,000 & 0.10 & 338.00\\

    \hline
  \end{tabular}
  \caption[Maximum energy consumption for scenarios]{Maximum energy consumption for scenarios. Scenarios described in Table~\ref{table:scenarios1}, with maximum energy consumed by all miners ($EnergyMax_{year}$), according to Equation~\ref{eq:energy-factors}. Values of maximum energy below 0 mean the corresponding scenario is not profitable.}
  \label{table:scenarios1-energy}
\end{table}

  Using formula~\ref{eq:energy-factors} we can estimate the maximum energy consumption of all Bitcoin miners per year for the scenarios defined in Table~\ref{table:scenarios1}. The result is shown in Table~\ref{table:scenarios1-energy}.
  
    It is interesting to notice that the value for the medium scenario should be close to the maximum energy consumption around May/June 2021, since the parameters of that scenario are close to those of these days.

\section{Impact of the different factors}
\label{sec:impact-factors}

Let's analyze the impact of these factors in some detail. As a part of that analysis, we will use data from the scenarios described in Table~\ref{table:scenarios1}.

\subsection{The impact of the price of energy}

Since we have $EnergyPriceEur$ in the denominator of all the fractions, it is clear that the lower the price of energy in traditional money (the way it is usually paid for), the higher the maximum energy that can be consumed by miners, still expecting a profit.

This fact could be expected, since in it just means ``all other factors equal, the cheaper the energy, the more miners will consume energy''.

For certain values of $Fees_T$, $AmortEUR$, $EurBTC$, we can define $NonEnergyBalanceEur$ (the balance of income and expenditure for non-energy factors, in Euro) as:

\begin{equation}
  NonEnergyBalanceEur_T = Blocks_T \times Mint \times EurBTC +
    Fees_T \times EurBTC - AmortEur_T
\end{equation}

And substituting in Equation~\ref{eq:energy-factors},

\begin{equation}
  EnergyMax_T = \frac{NonEnergyBalanceEur_T} {EnergyPriceEur}
\end{equation}

In other words, for a certain scenario of factors not directly dependent on energy consumption, the maximum energy consumed by all miners will be inversely proportional to the price of energy. The lower the price of the energy they can consume, the higher the total energy consumed.

Factors contributing to $NonEnergyBalanceEur_T$ are either constant, with the assumptions mentioned earlier in this paper, or independent factors: $Blocks_T$ and $Mint_T$ are constant. $Fees_T$ will depend on the number of transactions during $T$, $AmortEur_T$ will depend on amortization cost of equipment, and $EurBTC$ will depend on the conditions in the exchange market. Therefore, for some given non-energy factors, the lower the energy price miners can get, the higher their aggregated energy consumption.

\begin{figure}[ht]
  \centering
  \includegraphics[width=8cm]{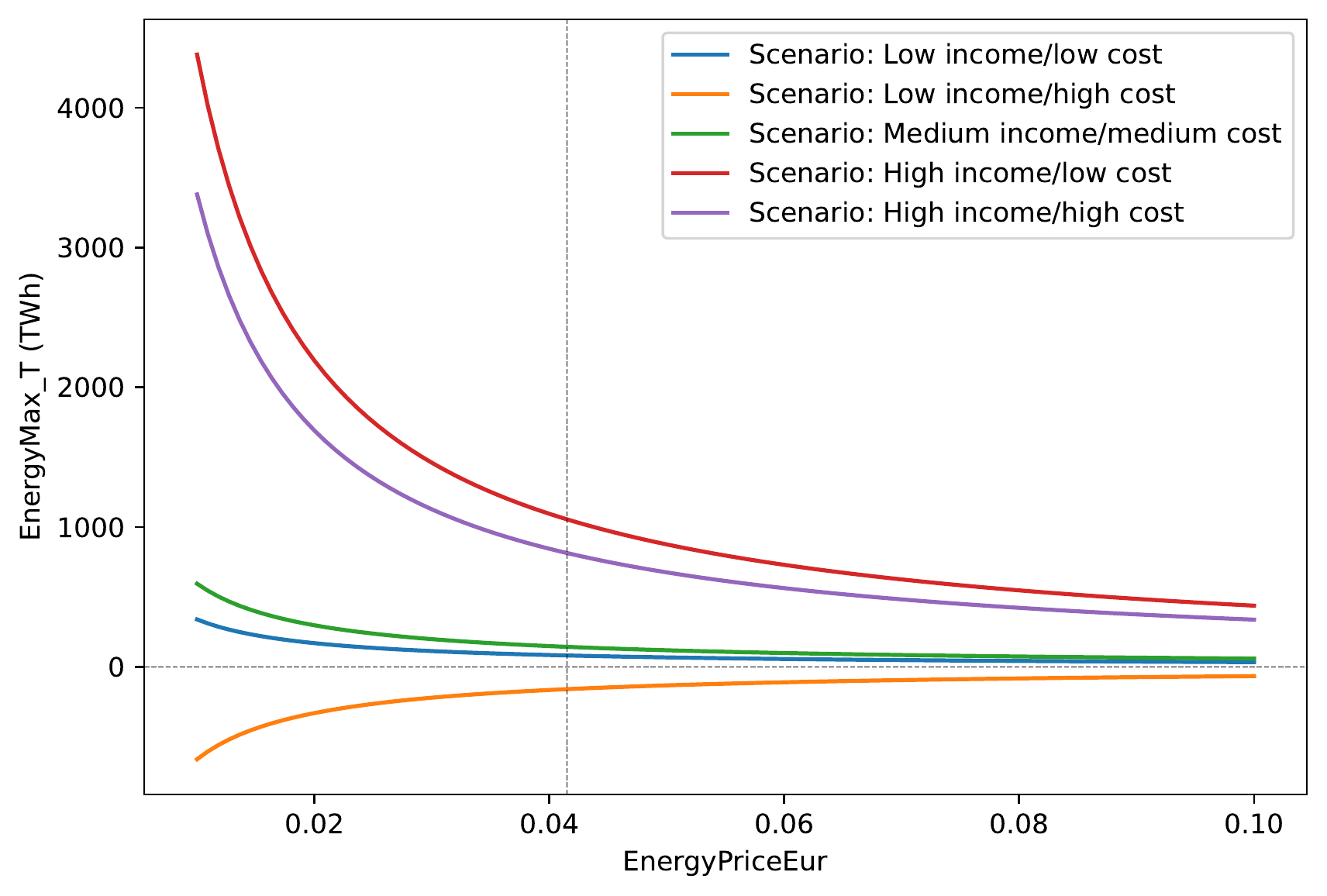}
  \caption[Impact of price of energy]{Impact of price of energy. Values of maximum energy consumed per year (in TWh) depending on the price of energy (in Euro per KWh), for Bitcoin exchange rate, amortization costs, and collected transaction fees of scenarios described in Table~\ref{table:scenarios1}. Values of maximum energy below 0 mean those prices of energy are not profitable for the corresponding scenario. The price for the vertical line corresponds to USD 0.05 per KWh, used by CBECI~\cite{cbeci_methodology:2021} as the average electricity price for miners.}
  \label{fig:energy-energypriceeur}
\end{figure}

Figure~\ref{fig:energy-energypriceeur} illustrates how the maximum energy consumed by Bitcoin depends on the price of electricity for scenarios defined in Table~\ref{table:scenarios1}.

\subsection{The impact of the exchange rate of Bitcoin}

The exchange rate of Bitcoin ($EurBTC$) is in the numerator of both fractions contributing to income. The income  obtained by miners in Bitcoin will therefore be proportional (in traditional money) to the exchange rate, which is easy to understand, since all income comes in Bitcoin. Therefore, an increase in the exchange rate for Bitcoin means, other things equal, more money to spend in electricity.

Defining the ratio of income in Bitcoin to energy price ($IncomeEnergyPriceRatio_T$) as

\begin{equation}
  Income_T =
    \frac{(Blocks_T \times Mint) + Fees_T} {EnergyPriceEur}
\end{equation}

maximum energy consumption is

\begin{equation}
  EnergyMax_T =
    \frac{Income_T}{EnergyPriceEur} \times EurBTC -
    \frac{AmortEur_T}{EnergyPriceEur}
\end{equation}

From this equation, the impact of the exchange rate is more apparent: for a given income in Bitcoin (minted plus fees), a given amortization cost, and a given energy price, the relation between energy consumption and the Bitcoin exchange rate is linear, with the following slope:

\begin{equation}
    \frac{Income_T}{EnergyPriceEur}
\end{equation}

\begin{figure}[ht]
  \centering
  \includegraphics[width=8cm]{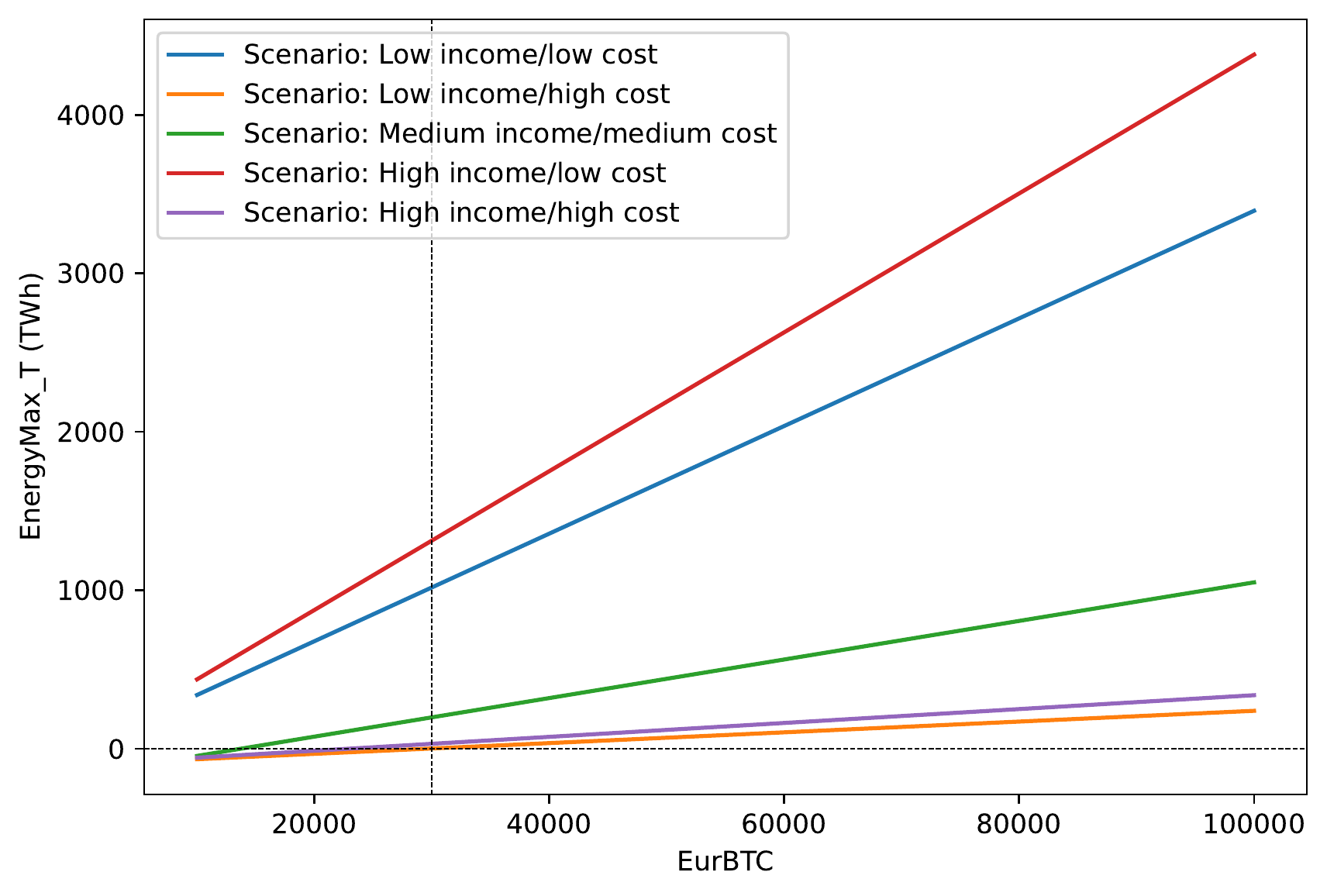}
  \caption[Impact of exchange rate]{Impact of exchange rate. Values of maximum energy consumed per year (in TWh) depending on the exchange rate of Bitcoin to Euro, for the price of energy, amortization costs, and and collected transaction fees of scenarios defined in Table~\ref{table:scenarios1}. Values of maximum energy below 0 mean those exchange rates are not profitable for the corresponding scenario. The exchange rate for the vertical line (30,000 Euro per Bitcoin) is close to the actual exchange rate for early June 2021.}
  \label{fig:energy-eurbtc}
\end{figure}

Therefore the, higher this ratio, the quicker $EnergyMAx_T$ will grow when EurBTC grows. This is shown in Figure~\ref{fig:energy-eurbtc}, which shows how the maximum energy that miners can consume varies with the exchange rate of Bitcoin to Euro, for two prices of energy (and the same $Income_T$).

Figure~\ref{fig:energy-eurbtc} illustrates how the maximum energy consumed per year depends on the Bitcoin exchange rate.

\subsection{The impact of minting}

The amount of minted Bitcoin per block is known in advance, and halves about every four years. This means that the number of minted Bitcoin that will be obtained by all the miners during period $T$ ($Mint$) also follows the same pattern. Therefore, in the short or medium term this factor has no impact on the energy consumption, but over time, since it will be decreasing, it will tend to drive energy consumption down.

Being minting one of the two incomes miners get, there is also a relationship with the other one, fees, which are discussed in the next subsection. For now, it is enough to say that, to maintain a certain income that balances a certain expenditure for all miners, when minting is halved, fees need to increase in the same quantity. If that does not happen, income will decrease, which will mean less expenditure, which other things being equal will mean less energy consumption.

\subsection{The impact of fees}

Fees ($Fees_T$) are in the numerator of an income fraction in~\ref{eq:energy-factors}. Therefore, the higher the fees collected by miners over period $T$, the more energy they can consume during that period, all other things equal.

Fees collected over a period are difficult to predict, but they are related to the number of Bitcoin transactions during the period. That means that, even when the relationship is not lineal, the more transactions during the period, the more energy can be consumed by miners, all other things equal.

\begin{figure}[h]
  \centering
  \includegraphics[width=8cm]{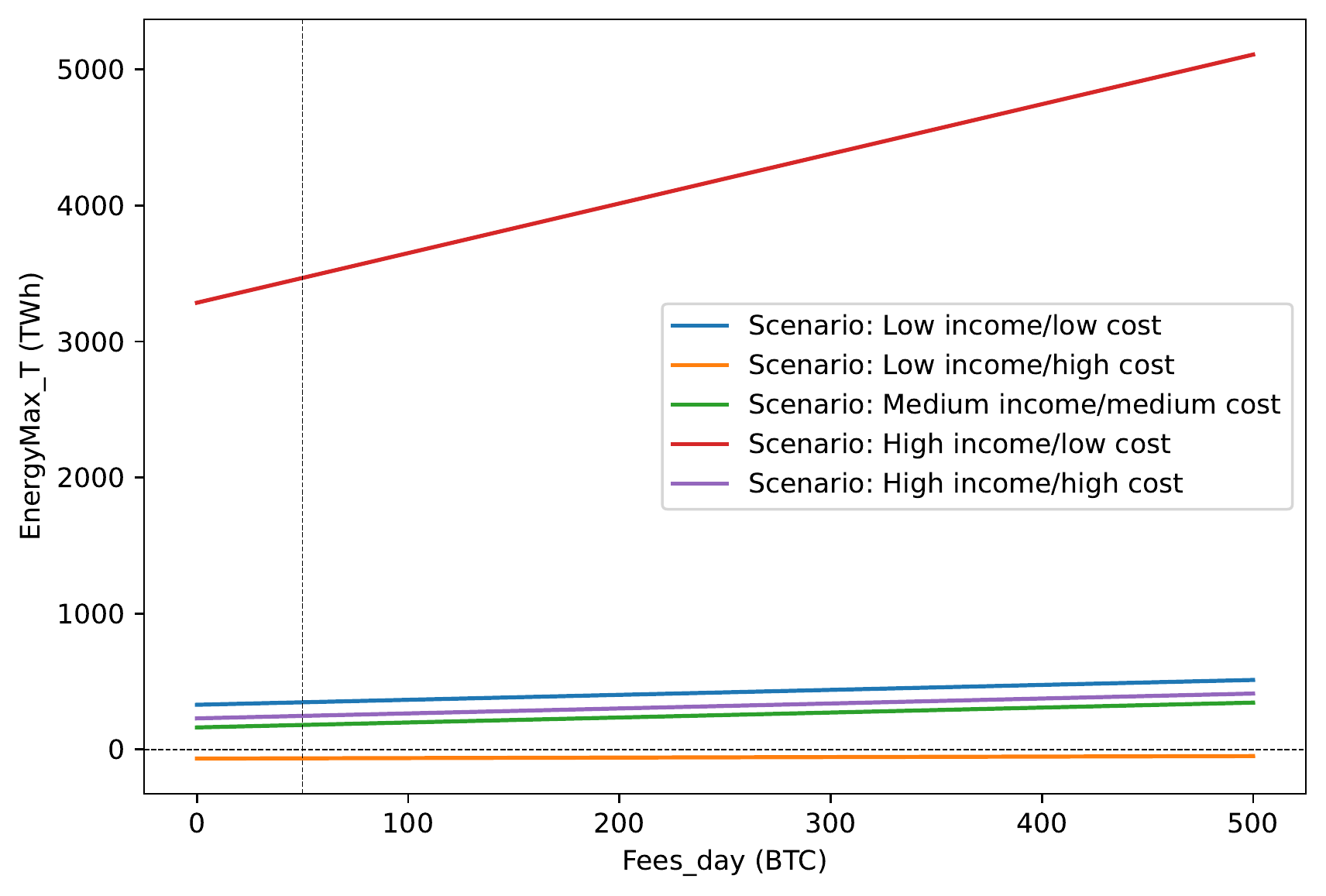}
  \caption[Impact of fees]{Impact of fees. Values of maximum energy consumed per year (in TWh) depending on fees collected from transactions, for the price of energy, amortization costs, and exchange rate of scenarios defined in Table~\ref{table:scenarios1}. Values of maximum energy below 0 mean those fees are not profitable for the corresponding scenario. Fees for the vertical line (50 BTC) is close to the weekly average of fees collected per day during early June 2021, according to Blockchain.com\protect\footnotemark.}
  \label{fig:energy-fees}
\end{figure}

\footnotetext{Blockchain.com (transaction fees): \url{https://www.blockchain.com/charts/transaction-fees}}

Figure~\ref{fig:energy-fees} illustrates the relationship of maximum energy consumed and transaction fees collected by miners for the five basic scenarios we are considering. 

An important consequence of the relationship the maximum energy consumed and the amount of fees is that it shows a control factor for the former. If fees can be lowered, the maximum amount of energy that miners can consume at profit is lowered too. Since fees tend to increase when the number of pending transactions for each block increases, lowering the number of transactions per unit of time for all Bitcoin would lower maximum energy consumption. Current approaches such as Lightning Network~\cite{poon2016:lighting_network} are proposing a second layer, on top of Bitcoin, that allows a large of number of transactions to be mapped on an single Bitcoin transactions. This way, the capacity of Bitcoin, in terms of transactions per unit of time, could increase without increasing the number of transactions in the Bitcoin blockchain, thus helping to keep fees and maximum energy consumption lower.

With the current ratio of fees to minted coins per block, this effect of lowering fees is relatively small. But in the future, as minting is reduced by halving, keeping fees under control will have a more noticeable impact on the maximum energy consumed by miners.

\subsection{The impact of amortization}

We have included in amortization costs ($AmortEur$) all the other costs (except for energy) that miners face to maintain their activity. According to the literature, the vast amount of it is amortization of the devices used for mining.
Since $AmortEur$ is in the numerator of the only negative addend in Equation~\ref{eq:energy-factors}, the relationship with the maximum energy consumption is linear, with a negative slope (the larger the amortization costs, the smaller the maximum energy consumption).

\begin{figure}[h]
  \centering
  \includegraphics[width=8cm]{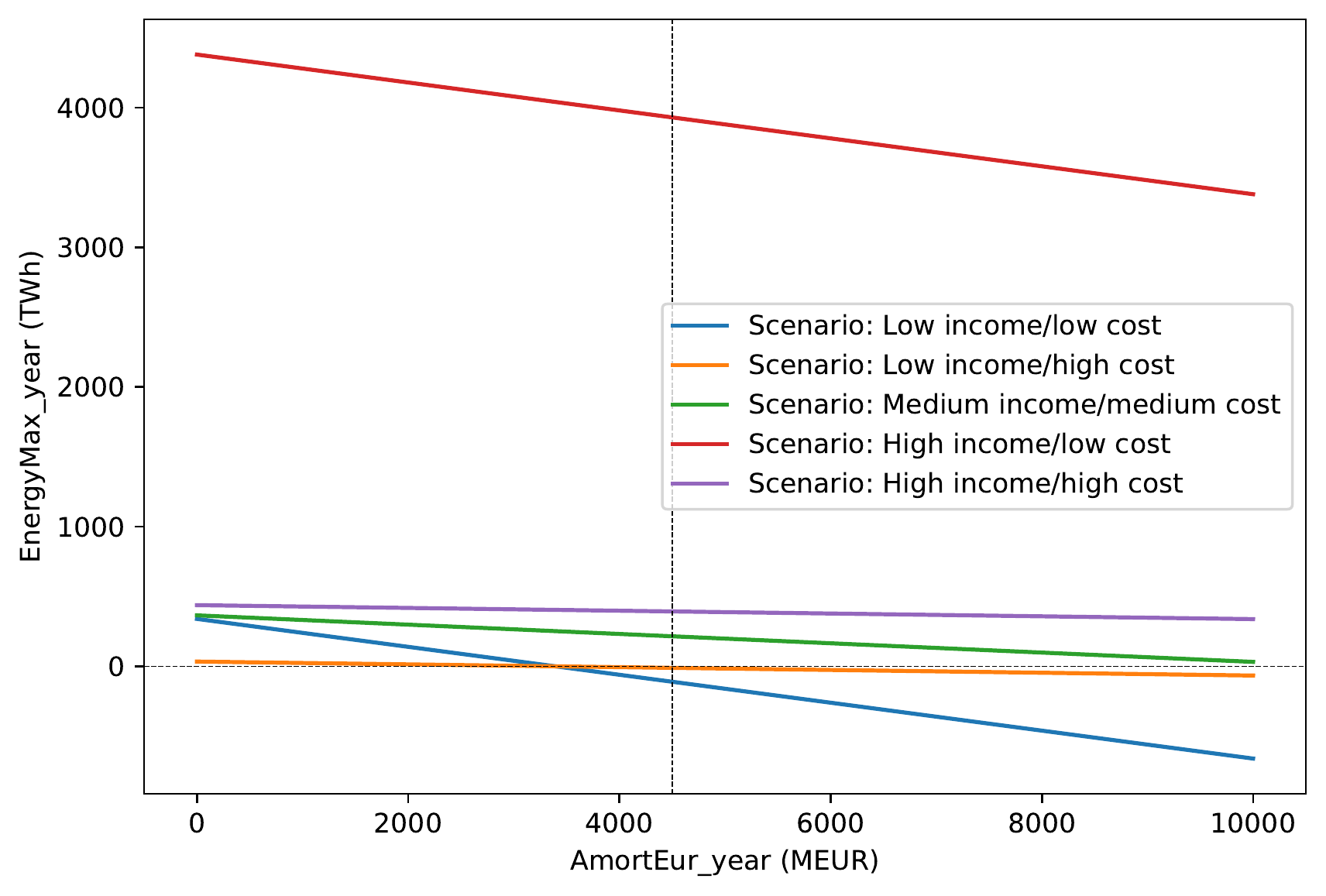}
  \caption[Impact of amortization]{Impact of amortization. Values of maximum energy consumed per year (in TWh) depending on amortization costs (in million Euro), for the price of energy, collected fees, and exchange rate of scenarios defined in Table~\ref{table:scenarios1}. Values of maximum energy below 0 mean those amortization costs are not profitable for the corresponding scenario. Amortization for the vertical line (4,500 MEUR) is estimated assuming miners are amortizing with prices and performance similar to that of the most efficient recent equipment (see Table~\ref{table:devices}).}
  \label{fig:energy-amort}
\end{figure}

In Figure~\ref{fig:energy-amort} we have depicted the relationship between amortization costs and maximum consumed energy for the scenarios we defined for this paper. In this figure we include a reasonable value for yearly amortization costs for all miners (vertical line, 4,500 MEUR/year), based on the values computed in Table~\ref{table:devices}.

The figure shows how the more miners invest in equipment (amortization), the less energy they consume. This may seem counterintuitive, because in the end more equipment will consume more energy. But in fact, more equipment only means miners could consume more energy, if they buy it. Miners will decide, based on the rest of the factors, if it makes sense to buy the equipment or not, if they want to stay profitable. If other factors (for example the price of energy or the exchange rate of Bitcoin) make buying more equipment non profitable, they will not buy it.

\subsection{The impact of the Amortization Factor}

The role of amortization in controlling energy consumption can be explored further if we realize that there is a relationship between amortization costs and energy consumed. That relationship is fixed by the price and energy consumption of mining devices. We can define the Amortization Factor, $AmF$, to capture this relationship, as the ratio between the amortization cost of a miner or a set of miners during a certain period $T$ and the energy consumed by them during that same period:

\begin{equation}
  AmF = \frac{AmortEur_T}{Energy_T}
\end{equation}

Using this factor, we can compute amortization costs for a certain period T, for a certain set of miners with $AmF$ as their Amortization Factor, and consuming $Energy_T$ as:

\begin{equation}
  AmortEur_T = Energy_T \times AmF
\end{equation}

With this, we can rewrite Equation~\ref{eq:cost-income-equilibrium}, expressing both sides in Euro, as:

\begin{equation}
  (Energy_T \times AmF) + (Energy_T \times EnergyPriceEur) \leq
  ((Blocks_T \times Mint) + Fees_T) \times EurBTC
  \label{eq:cost-income-amf-equilibrium}
\end{equation}

And finally, we get the maximum aggregated energy consumption of all bitcoin Miners, by applying the inequality to all Bitcoin miners in the expenditure-income equilibrium:

\begin{equation}
  Energy_T =
  \frac{((Blocks_T \times Mint) + Fees_T) \times EurBTC}{AmF + EnergyPriceEur}
\end{equation}

The numerator in this fraction is the aggregated income, in Euro. So, we can simplify the equation to help understand its consequences, obtaining the ``Bitcoin Energy Amortization Factor Formula'' for Bitcoin miners aggregated energy consumption:

\begin{equation}
  Energy_T =
  \frac{IncomeEur}{AmF + EnergyPriceEur}
  \label{eq:energy-income-amf}
\end{equation}

This shows clearly how there are two factors that control energy consumption, given a certain income in euros (which is controlled by Bitcoin obtained per block, and by the Bitcoin exchange rate). These two factors are the price of energy in Euro, which we already explored, and the Amortization Factor. It is interesting to notice that, when $AmF$ is much lower than $EnergyPriceEur$, energy consumption is almost controlled by $EnergyPriceEur$. But if the Amortization Factor is large enough, it is necessary to have it into account. In fact, if it becomes much larger than $EnergyPriceEur$, it will control almost alone the total energy consumed by Bitcoin miners.

\begin{table}
  \begin{tabular}{|r||r|r|r|}
    \hline
    Name & $AmortEur_T$ & $Energy_T$ & $AmF$  \\
         & (EUR)  &  (KWh)  & (EUR/KWh) \\ \hline\hline
    Antminer S19j Pro (100 Th) & 7,080 & 53,436 & 0.13 \\
Antminer S19j Pro (110 Th) & 8,797 & 56,940 & 0.15 \\
Antminer S9 SE (16Th) & 829 & 22,426 & 0.04 \\

    \hline
  \end{tabular}
  \caption[Amortization factors]{Amortization factors of devices in Table~\ref{table:devices}. We consider devices are fully amortized in two years, and use a period for the calculus equal to that amortization period ($T = 2 \times 365 \times 24$, in hours).}
  \label{table:devices-amf}
\end{table}

Table~\ref{table:devices-amf} shows the values of Amortization Factor for the devices presented in Table~\ref{table:devices}. It is clear that, for all devices considered, $AmF$ is comparable to the cost of energy in all our scenarios. In fact, for the most energy-efficient equipment, for low prices of energy (for example, 0.01 Euro/KWh) the impact of $AmF$ in the aggregated energy consumed by all miners is more than 10 times that of the price of energy.

\begin{figure}[h]
  \centering
  \includegraphics[width=8cm]{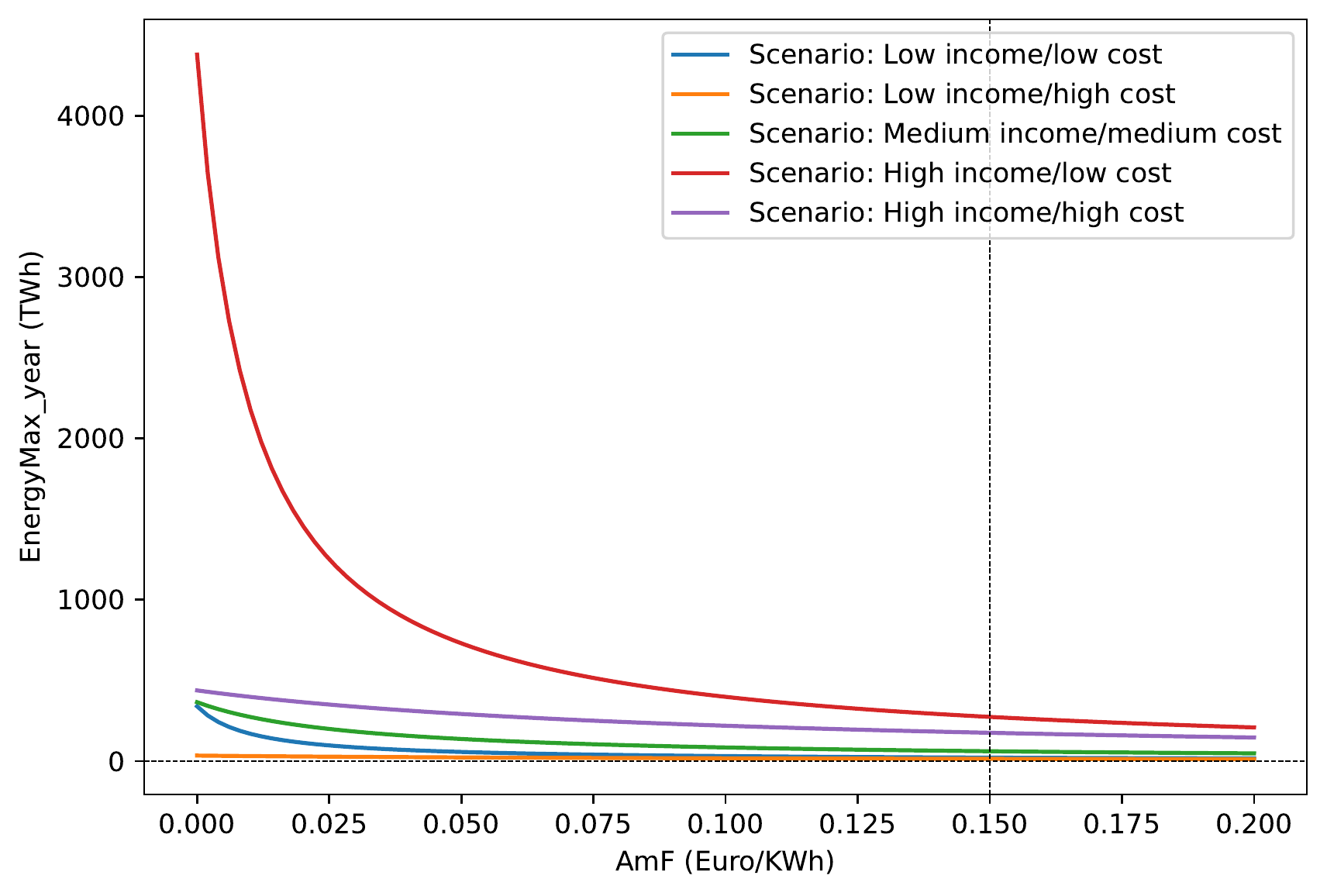}
  \caption[Impact of Amortization Factor]{Impact of Amortization Factor. Values of maximum energy consumed per year (in TWh) depending on the value of Amortization Factor for the price of energy, collected fees, and exchange rate of scenarios defined in Table~\ref{table:scenarios1}. The vertical line corresponds to the Amortization Factor of the most recent device in  Table~\ref{table:devices} (0.15)}
  \label{fig:energy-amf}
\end{figure}

Figure~\ref{fig:energy-amf} shows how the total energy consumption varies with the Amortization Factor, for the different scenarios defined earlier. One of the most interesting takeaways of this chart is how, above a certain threshold, all the lines become close to flat, and total energy consumed is relatively low.

\subsection{The impact of aggregated hashpower}
\label{sec:impact-hashpower}

The aggregated hashpower of all miners has a direct influence in the energy that miners consume:

\begin{equation}
  EnergyEur_T = DevicePower \times \frac{HashPower}{HashPowerDevice} \times T
\end{equation}

However, the aggregated hashpower is only a consequence of the income miners get. This is the reason why the aggregated hashpower does not appear in Equation~\ref{eq:energy-factors}. In fact, the hashing power needed to produce a block is adjusted (approximately) every week by the Bitcoin protocol. If other factors allow miners to add hashpower by putting in operation new devices, the difficulty of mining a block will be increased in less than a week, rendering the increase void with respect to collecting (collectively) more income: they can only collect the aggregated income produced.

In the equilibrium, miners will tune hashpower, by switching on or off devices, according to how much money they can spend to balance their income. We can define cost per hash in Euro ($CostHashEur$), which is characteristic of each device, given a certain price for energy: it will depend on its hashpower and its energy consumption. The aggregated hashpower can be computed as:

\begin{equation}
  HashPower = \frac{IncomeEur_t}{CostHashEur \times T}
\end{equation}

This leads to the interesting result that even when individual miners compete in hashpower, since the larger the fraction of hashpower they have, the larger the income), for all of them together this is a zero-sum game: their total income is independent of the aggregated hashpower of all miners. In fact, it is worse than zero-sum, because the more hashpower, the more costs (in the form of energy and amortization costs), which means that they are competing for the same income with increased costs. If miners could control the total hashpower, for them it would be more profitable to keep it as low as possible, since that is a cost for all of them (both in terms of energy and amortization). This would mean that all miners together would stay way below their maximum energy consumption, to minimize their aggregate costs. But as long as joining the set of miners remain a competitive action, the total hashpower will tend to be as high as possible, since as long as there is profit, every single miner will tend to invest as much as possible in both energy and amortization, or new miners will.

We can also see how, in the end, the aggregated hashpower will tend to follow the exchange rate of Bitcoin since, from Equation~\ref{eq:income}, we can substitute $IncomeEur_T$ for its factors: 
\begin{equation}
  HashPower = \frac{(Blocks_T \times Mint) + Fees_T}{CostHashEur \times T} \times EurBTC
  \label{eq:hashpower-income-cost}
\end{equation}

This means that, for some given block minting and transaction fees, the aggregated hashpower will be proportional to the exchange rate of Bitcoin. Which is consistent with the discussion we already had on the impact of the exchange rate on energy consumption.

\section{Discussion and threats to validity}
\label{sec:discussion}

After exploring the different factors that affect aggregated energy consumption of all miners, let's now discuss these factors, which of them could be used to control, to some extent, energy consumption, and the validity of the main assumptions that base our exploration.

\subsection{Factors controlling energy consumption}

Of the many factors controlling energy consumption of Bitcoin miners, there are two that historically have been more volatile, and difficult to predict:

\begin{description}
\item[Exchange rate] Bitcoin miners receive all their income in Bitcoin. Therefore, while their expenditure is mainly in traditional currency, the exchange of Bitcoin to traditional currency is a major driver of their aggregated income. Since this exchange has been very volatile in the past, their income is also very volatile. In particular, in periods of a rising Bitcoin, miners income increases proportionally, and as we saw, their energy consumption too (being other factors constant). This causes problems for miners in periods of a declining exchange rate, since their income decreases, and they can no longer afford amortization and energy costs. Since they can only really act on energy costs (they need to face amortization costs whatever happens), it would seem rational for them to reduce energy consumption more than equations of equilibrium suggest. However, the entry of new miners not subject to amortization cost of past equipment, could mitigate this mechanism.

  In any case, this factor alone may explain the great increase in energy consumption during the past years, since the exchange rate of Bitcoin has increased a big amount during this period. The volatility of the exchange rate also means that miners cannot act based only on current or past exchange rates: they need to predict the exchange rate during the next months, or years, to adequately plan for their purchases and mining operations. Therefore, it is likely that it is more the expected exchange rate than the current exchange rate what drives their decisions. In any case, in the long term, the equilibrium presented in this paper will hold. If miners overspend, buying more equipment and consuming more energy than the income they get, sooner or later they will go bankrupt, and be substituted by others who can adjust their costs better.

  Since the exchange rate of Bitcoin is fixed in the market, based on its perceived utility and other factors, it can be considered to be out of control of any agent, and therefore cannot be used to control aggregated energy consumption.
  
\item[Transaction fees] Transaction fees is the volatile part of miners income in Bitcoin. Minted coins also change, but they change only every four years, and in a predictable way. Income via transaction fees are dependent on the fees users of Bitcoin are ready to face, which in turn are related to the size of the queue of pending transactions, and the urgency of users to have their transactions included in the blockchain.

  With time, as minted coins per block get halved every four years, fees are expected to be the major source of income for miners. Therefore, they are already an important driver of energy consumption, and will be even more important with time. If pending transactions, and their urgency to enter the blockchain, are reduced, transaction fees will likely be reduced too.
\end{description}

Other variable, but more stable factors affecting energy consumption are:

\begin{description}
\item[Minting] This mechanism is entirely predictable: minted coins per block will be halved every approximately four years. This means that income for miners will also be reduced every four years, and that fact should be reflected in the energy they can consume (being other factors equal).

\item[Energy price] The price of energy is given by the technologies available to produce the energy that miners consume, and other factors such as taxing. The current evolution of the cost of energy production is clearly directed towards decreasing costs. Other factors are more difficult to predict. However, given that miners can relocate with relative ease, it is expected that they move as much as possible to places with access to cheap energy, and in general their energy costs will be close to the lower costs that technology and other factors allow.

\item[Amortization cost] Cost of equipment, and its duration, are the main factors that affect amortization cost. Cost per hash is in general declining, as technology improves. But as we showed, the real impact on energy consumption is in the relationship between equipment cost and duration (amortization) and energy consumed. During the last years, it seems that this ratio (Amortization Factor) is increasing, with a higher amortization cost per unit of energy consumed. This evolution is reducing aggregated energy consumption, being other factors equal. Likely the reason for this trend is the increasing specialization of mining equipment, from general purpose CPUs to GPUs to ASICs.
\end{description}

\subsection{Actions that could decrease energy consumption}

Some specific actions that influence some of the factors could lead to reducing energy consumption by Bitcoin miners, not impacting the chances or controlling the aggregated hashpower (leading to majority attacks on the blockchain), neither interfering with the fundamentals of the proof of work mechanism. The two actions that are most remarkable in this respect are tuning the Amortization Factor of mining devices, and reducing transaction fees:

\begin{description}
\item[Tuning the Amortization Factor] The Amortization Factor shows possible scenarios in which more expensive machines, with less energy consumption, do not impact the aggregated hashpower neither the costs that miners have to face to mine blocks, but have a lower aggregated energy consumption. In some sense, the Amortization Factor, and the relationship of income and expenditure for all miners together could lead to consider ``proof of cost'' as an extended version of the traditional ``proof of work''. In the end, to produce its benefits, avoiding concentration and minimizing majority attacks, Bitcoin needs that a high cost is distributed among a large number of independent miners. Consuming energy is in fact a sizable part of this cost. But when the Amortization Factor of equipment used to mine is high, amortization is a much larger cost than energy consumption. Thus, ``proof of cost'', tailored to minimize energy consumption, may have all the benefits of ``proof of work'', but minimizing the problems of energy consumption. To which extent Amortization Factor will increase in the future will depend on the technological evolution of mining devices, but also could be influenced by tuning of the specific algorithms used to produce hashes.

  This is just a consequence of Equation~\ref{eq:energy-factors}, where it is clear that, on the expenses side, every euro spent in equipment is an euro not spent in energy. The Amortization Factor, characteristic of each device, gives a tool to better understand this effect, as shown in Equation~\ref{eq:energy-income-amf}.

\item[Reducing transaction fees] Transaction fees depend on how many transactions are performed, and how quickly people want them to be included in the blockchain. Therefore, any mechanism that influence the number of transactions, or the need to have them quickly included in the blockchain, has the potential of reducing transaction fees. Mechanism that build transaction layers on top of Bitcoin (off-chain transactions) usually help in both directions. Because of their nature, the sort of aggregate many higher layer transactions into a few Bitcoin transactions, therefore making it possible to have many transactions between users with a much lower number of ``real'' Bitcoin transactions. Usually, since the upper layer is good enough for most uses, ``real'' Bitcoin transactions that reflect groups of upper layer transactions can also wait for longer periods until they are included in the blockchain. Therefore, these mechanisms tend to reduce transaction fees at the Bitcoin level (even if they may include other fees at higher levels), which make miners income to be lower, which reduces the energy they may consume in the profit equilibrium.

\item[Accelerating the halving] ``Halving'' is the process of reducing the coins minted per block. Currently, Bitcoin is halving minted coins every approximately four years. If this process would accelerate, say to halving every two years, income per minting will be dramatically reduced in a few years, with the consequent impact in energy consumption. This acceleration would not have a significant impact on the amount of Bitcoin in circulation (most coins were already minted), neither in the total number of Bitcoin that will be produced (will be the same, only it will be reached earlier). It also doesn't show an impact on the concentration of miners.

\item[Increasing energy price] Given that miners relocate relatively easily, regulating the price of energy for them is not easy, except if worldwide regulations are enforced. However, if the needed agreements could be reached, the price of energy for miners could be increased, for example by taxing, thus reducing the aggregated consumption. However, since usually it is not exactly energy consumption what may be a problem, but energy consumption with an impact on the environment, access of miners to non-clean energy could be controlled or made more expensive, thus acting not on the total energy consumed, but on the non-clean energy consumed, which could be appropriate in some cases.
\end{description}

\subsection{Validity of the main assumptions}

Some assumptions that influence the results presented in this paper could be wrong, therefore affecting some parts of the analysis:

\begin{description}
\item[Miners are operating at income / expenses equilibrium]
  
Our exploration of the factors that determine the maximum energy consumption by Bitcoin miners is based on the analysis of the equilibrium between their income and their expenditures. However, the assumption that miners operate in this equilibrium is just a convenient artifact. Usually, miners will tend to work for profit, which means that they will try to cut expenditures as much as possible, and their benefit (the difference between their income and their expenses) should also be accounted.

However, since miners are competitive agents, the moment one of them find ways to reduce expenses while obtaining the same income others will try to do the same, nullifying their advantage, and tending to an aggregated zero profit. This will be reflected in general in the cost of a hash being equal to the income for producing a hash: miners that can reduce the cost per hash will tend to add more hashpower, to maximize their expected income and benefits. That will expel from the set of miners to those who are less efficient, and with a higher cost per hash cannot afford to compete. Usually, they will be substituted by others, more efficient miners, and benefits will come again close to zero. In the past, we have seen this once and again, with the substitution of miners using traditional CPUs by miners using GPUs, then with miners using ASICs. Or with miners finding increasingly cheaper sources of energy.

Therefore, although in the study we have been interested in the maximum energy consumed, the numbers found should be, if competition is really working, quite close to the actual numbers for energy consumption. In other words, Bitcoin miners are likely working very close, as a whole, to the economic equilibrium of profits and expenses. Of course, any deviation from this assumption will mean deviations in the actual energy consumed.

\item[Miners are competing with other miners]

  In general, miners are considered to be competing agents, trying to get as much benefit as possible without other agreements with other miners than those imposed by the Bitcoin functioning. However, if miners could collude, and somehow find ways to reduce the aggregated hashpower below what Equation~\ref{eq:hashpower-income-cost} mandates, they could collectively have more benefits, and consume less energy. In other words, if competing among themselves, miners are forced to operate as much hashpower as permitted by the equilibrium equation, spending all their income in equipment and energy. But if they could agree on an upper threshold to hashpower, somehow impeding other new miners to fill the gap until the theoretic maximum, they could devote a part of their income to profit, instead of expenses.

  It is very unlikely that, given how easy it is to become a miner, and how distributed is the population of miners, these kind of thresholds can be imposed, so in general we assume this assumption holds.
  
\item[Energy consumption corresponds to energy consumed by mining devices]

  In real mining facilities, energy is not needed only to power mining devices. It is used also for cooling, maintenance, and other tasks. When we analyze the aggregated balance of expenses for miners, this is not a problem, since ``energy consumed'' is just a cost, and can refer to any kind of energy needed by miners. However, in the analysis of the Amortization Factor, the nominal energy consumption of devices is used. To be more real, the analysis should also include other kinds of energy needed to mine. In general, they could be computed as a multiplier of the nominal energy consumption, to get more realistic values for the Amortization Factor of a miner

\item[Amortization cost is derived only from the cost of devices]

  Miners will usually face other amortization costs, not derived from the cost of purchasing mining equipment. They will have installation costs, infrastructure costs, management costs, etc. When we analyze the aggregated balance of expenses for miners, this is not a problem, since ``amortization'' is just a cost, reflecting all costs that are not energy. However, in the analysis of the Amortization Factor, the analysis should include other non-energy costs. As in the previous case, those costs could be computed as a multiplier of amortization costs.

\item[The period of amortization is known in advance]

  Our analysis analysis assumes that the period of full amortization (the period from the moment some mining device is put in operation to the moment it is shut down because it is no longer profitable) is known in advance, so that miners can have it into account to make decisions. However, this is usually not the case. Equipment break-down, and variations in Bitcoin exchange rate and energy cost may suddenly make some equipment profitable or unprofitable during some periods of time. These events are not predictable, and therefore, miners can only work with assumptions about the future.

  This does not affect the general analysis, since whatever their assumption, either they are close to reality and then they succeed, or not and they will go bankrupt and be substituted by other miners. However, for the specific analysis of some situation, and therefore the aggregated energy consumption for it, those assumptions matter. For example, if the current exchange rate of Bitcoin is 20,000 Euro/BTC, but miners expect it to be 100,000 Euro/BTC in six months, they will try to acquire equipment to be ready for that new exchange rate.

  Therefore, although energy consumption computed for long periods should be close to reality, for short periods it could be quite different, especially if miners have expectations of significant changes of scenario in the near future.

\item[Miners operate with the most profitable equipment]

  Again, for the general analysis this is not relevant. But for estimating the energy consumption in a given scenario, our analysis assumes that miners are operating with the most profitable equipment available. In general, this will not be true: miners will operate with the equipment they already have. But in general, the assumption will hold for the new equipment they purchase, since they want to maximize their income. Therefore, as time passes, the installed base of devices gets renovated, as they become unprofitable, and the new equipment fits the assumption.

  Therefore even when the assumption does not hold in reality, we assume that reality is close enough to it, thanks to this renovation mechanism, to be useful to compute energy consumption.

\item[Fees are known in advance]

  Once again, for the general analysis this assumption is not relevant. But for miners to make their decisions, income is as important as expenditure. Income from minting is completely predictable (in Bitcoin), but fees are not. They depend on the queue of transactions pending entering the blockchain, and on expectations by Bitcoin users.
\end{description}

\section{Related work}
\label{sec:related-work}

There have been many studies estimating the energy consumption of Bitcoin miners. In general, they start from estimating the aggregated hashpower of all miners, and then, based on the hashpower of most common hashing (mining) devices in use, estimate the number of total hashing devices (miners). With this, they can just multiply by the energy consumption of these devices to get the aggregated consumption for all miners. A good example of this approach is described in~\cite{bevand17:_elect_bitcoin}, which summarizes its method as follows:

\begin{quote}
  ``I drew a chart juxtaposing the Bitcoin hash rate with the market availability of mining ASICs and their energy efficiency. Using pessimistic and optimistic assumptions (miners using either the least or the most efficient ASICs) we can calculate the upper and lower bounds for the global electricity consumption of miners''
\end{quote}

This method is used by the Cambridge Bitcoin Electricity Consumption Index (CBECI)\footnote{CBECI: \url{https://cbeci.org/}}, which estimates an aggregated energy consumption of all miners of 90.10~TWh, with a range between 33.66~TWh and 226.24 TWh, both for June 15th 2021.

Even when this approach is quite helpful to estimate the total energy consumption, based on a detailed model of which equipment miners are using, and its individual energy consumption, it is not very useful to reason about the influence of other factors (such as the Bitcoin exchange rate, or the price of electricity) on the quantity and quality of the equipment actually in use. In other words, although this strategy is very useful to estimate the current aggregate consumption, it is not that useful for reasoning about the factors which ultimately drive that consumption.

In this paper we have followed another approach, based on assuming that miners will in general operate at profit, and studying the factors that influence that profit. To our knowledge, this approach was first explored in \cite{RePEc:new:wpaper:1505}, which focused on the estimation of the cost of ``production'' of Bitcoin as a predictor of its exchange rate:

\begin{quote}
  ``Bitcoin production seems to resemble a   competitive
market, so in theory miners will produce until their marginal
costs   equal   their   marginal   product.   Break-even   points   are
modeled for market price, energy cost, efficiency and difficulty to
produce.   The   cost   of   production   price   may   represent   a
theoretical value around which market prices tend to gravitate.''
\end{quote}

In the process of estimating this ``production cost'', energy consumption is estimated, based on the equilibrium between the cost of energy consumed (expenditure) and the exchange value of Bitcoin produced by minting (income) per day, for a given hashpower of a miner. Then the cost that a miner faces for minting a Bitcoin is considered a lower bound for the exchange rate of Bitcoin, as it would be the ``production cost'' of that Bitcoin. Although the paper does not provide an specific equation or estimation value for the aggregated energy consumption of all miners, it is straightforward to extrapolate it from equations shown in the paper. For the estimation of the cost, the paper does not considers amortization of equipment or any other cost except for energy consumption, and for income it does not consider transaction fees.

There is also in this paper an interesting discussion on how, below a certain exchange rate, miners would shutdown equipment which is not profitable for them because it consumes too much energy. But the analysis does not considers what happens if for any reason the exchange rate is kept at that value: aggregated hashpower will drop because of the retiring miners, the difficulty of the Bitcoin protocol will drop, to adjust the number of Bitcoin minted per period of time, and the remaining miners will be profitable again because minted Bitcoin per hash will therefore increase. This would lead to a new stable situation, with a lower aggregated hashpower, and therefore a lower energy consumption, not altering the exchange rate of Bitcoin. In other words, as we showed, the aggregated hashpower will adjust to the other factors actually driving energy consumption.

This approach is further explored in~\cite{vries2017:bitcoin_elect_consum}\footnote{A peer reviewed version has been published as \cite{vries21:_bitcoin}, but I couldn't access it.}, where it is specifically used to estimate an upper bound for aggregated energy consumption. However, this paper does no have into account transaction fees, and only in a very limited way has into account amortization costs. It is also more focused in finding upper and lower levels for energy consumption given a certain scenario than in exploring the interrelationship between the different parameters influencing energy consumption.

This approach is also the basis of the Bitcoin Energy Consumption Index (BECI)\footnote{BECI: \url{https://digiconomist.net/bitcoin-energy-consumption/}}, which estimates a minimum consumption per year, as of June 15th 2021, of 42.22 TWh, and an estimated consumption per year of 121.29 TWh (both as of June 15th 2021).

\section{Conclusions}
\label{sec:conclusions}

In this paper we have explored the factors determining determining the maximum energy consumption by Bitcoin miners, based on an analysis of the equilibrium between costs and expenditures for miners. Some of the main results of this exploration are:

\begin{itemize}
\item A method, based on the analysis of income and expenses of all miners, to find the economic equilibrium in which they should operate if they are on the verge of being profitable, and at the same time competing.
\item The Bitcoin Energy Factors Formula, Equation~\ref{eq:energy-factors}, which relates energy consumption to Bitcoin minting, transaction fees, energy price, amortization cost, and Bitcoin exchange rate. This equation shows how all these factors impact the aggregated energy consumption of all miners.
\item The Bitcoin Energy Amortization Factor Formula, Equation~\ref{eq:energy-income-amf}, which relates energy consumption to the aggregated income of all miners, the Amortization Factor (a ratio characterizing mining equipment), and the price of energy.
\item A detailed explanation of how individual factors affect aggregated energy consumption.
\item An analysis of some scenarios, characterized by the values of the different factors, computing the aggregated energy consumed in each of them.
\item A proposal of some mechanisms that could be used to control the aggregated energy consumption.
\end{itemize}

Not necessarily all of these contributions are new, but they helped me to make up my mind, and to reason about how energy consumption of Bitcoin miners actually works, beyond just estimating how big it is, and inferring how it will be based on current trends. Instead of that, looking at the fundamental factors that affect it, I think we can better reason about its future trends, and even start discussions about how it could be controlled, while at the same time maintaining the main characteristics of the current Bitcoin protocol.

\textbf{Reproduction package} There is a GitLab repository\footnote{GitLab repository for this paper: \url{https://gitlab.com/jgbarah/bitcoin-paper}} with source for this paper, including a Jupyter Python notebook and some CSV files with data in scenarios. The notebook can also be interactively consulted in Binder\footnote{Notebook in Binder: \url{https://mybinder.org/v2/gl/jgbarah\%2Fbitcoin-paper/HEAD?filepath=notebooks\%2Fanalysis.ipynb}}.

\textbf{Feedback about this paper} Please provide feedback about this paper as you may think it will reach me. This said, opening an issue in GitLab\footnote{Feedback: \url{https://gitlab.com/jgbarah/bitcoin-paper/-/issues/new}} is the preferred feedback method.

\printbibliography

\end{document}